\documentclass[sigconf]{acmart}
\acmSubmissionID{14}

\AtBeginDocument{%
  \providecommand\BibTeX{{%
    \normalfont B\kern-0.5em{\scshape i\kern-0.25em b}\kern-0.8em\TeX}}}

\begin{document}

\makeatletter
\renewcommand\@formatdoi[1]{\ignorespaces}
\makeatother

\acmYear{2020}\copyrightyear{2020}
\setcopyright{acmcopyright}
\acmConference[ACM CHIL '20]{ACM Conference on Health, Inference, and Learning}{April 2--4, 2020}{Toronto, ON, Canada}
\acmBooktitle{ACM Conference on Health, Inference, and Learning (ACM CHIL '20), April 2--4, 2020, Toronto, ON, Canada}
\acmPrice{}
\acmDOI{10.1145/3368555.3384449}
\acmISBN{978-1-4503-7046-2/20/04}
\title{Disease State Prediction From Single-Cell Data Using Graph Attention Networks}

\author{Neal G. Ravindra}
\affiliation{%
  \institution{Yale University}
  }
\authornote{Both authors contributed equally to this research.}
\authornote{Internal Medicine(Cardiology), Computer Science}
\email{neal.ravindra@yale.edu}

\author{Arijit Sehanobish}

\affiliation{%
  \institution{Yale University}
  }
\authornotemark[1]
\authornotemark[2]
\email{arijit.sehanobish@yale.edu}

\author{Jenna L. Pappalardo}
\affiliation{%
  \institution{Yale University}
  }
 \authornote{Neurology, Immunobiology}
\email{jenna.pappalardo@yale.edu}

\author{David A. Hafler}
\affiliation{%
  \institution{Yale University}
  }
  \authornotemark[3]
  \email{david.hafler@yale.edu}

\author{David van Dijk}
\affiliation{%
  \institution{Yale University}
  }
  \authornotemark[2]
 \email{david.vandijk@yale.edu}

\renewcommand{\shortauthors}{Ravindra et al.}

\begin{abstract}
Single-cell RNA sequencing (scRNA-seq) has revolutionized biological discovery, providing an unbiased picture of cellular heterogeneity in tissues. While scRNA-seq has been used extensively to provide insight into both healthy systems and diseases, it has not been used for disease prediction or diagnostics. Graph Attention Networks (GAT) have proven to be versatile for a wide range of tasks by learning from both original features and graph structures. Here we present a graph attention model for predicting disease state from single-cell data on a large dataset of Multiple Sclerosis (MS) patients. MS is a disease of the central nervous system that can be difficult to diagnose. We train our model on single-cell data obtained from blood and cerebrospinal fluid (CSF) for a cohort of seven MS patients and six healthy adults (HA), resulting in 66,667 individual cells. We achieve $\mathbf{92}$ \% accuracy in predicting MS, outperforming other state-of-the-art methods such as a graph convolutional network and a random forest classifier. Further, we use the learned graph attention model to get insight into the features (cell types and genes) that are important for this prediction. The graph attention model also allow us to infer a new feature space for the cells that emphasizes the differences between the two conditions. Finally we use the attention weights to learn a new low-dimensional embedding that can be visualized. To the best of our knowledge, this is the first effort to use graph attention, and deep learning in general, to predict disease state from single-cell data. We envision applying this method to single-cell data for other diseases.

%

\end{abstract}



\maketitle
\fancyfoot{}

\section{Introduction}

Single cell RNA-sequencing (scRNA-seq) technology provides gene expression data for individual cells, yielding omics-scale information at single-cell resolution. The information contained within transcriptomes of individual cells can be used to identify rare cells, genes, and other changes associated with fundamental biological processes and pathological states. However, identifying hidden differences in cell pathophysiological trajectories is challenging. ScRNA-seq data typically contains gene count data for more than $20,000$ transcripts across $10,000$ - $100,000$ cells and has a high-degree of technical noise, and encompasses a large degree of biological variability \cite{compchallenge, rnacluster}. While these data promise to provide rich insight into complex systems that determine health and well-being, integrative computational tools are required to comprehensively learn about cell states from scRNA-seq data collected across cell, disease, and temporal modalities \cite{rnaanalysis}. 

Computational approaches have been developed to use scRNA-seq data to identify bio-markers for potential use in diagnostics \cite{Gawel2019}, to understand genetic factors underlying disease heterogeneity \cite{ZengAndDai2019}, and to predict cellular response to experimental perturbations \cite{Lotfollahi2019}. In addition, several classifiers, using primarily non-deep learning methods, have been employed to predict cell-types from single-cell samples \cite{cellpred, Zheng2019}. However, to the best of our knowledge, no computational approaches have been developed to harness scRNA-seq data to predict disease states for the use of transcriptomic technology in diagnosis \cite{rnadiagnosis}. Despite this gap, scRNA-seq data may revolutionize our understanding of the biological processes associated with various disease states in addition to providing personalized insight into a patient's pathophysiological state. These kinds of omics-scale insights may be useful in tailoring treatments for personalized medicine \cite{omics}. We propose a learning approach that harnesses the rich-information content in a scRNA-seq dataset to characterize and predict two pathological states related to neuroinflammation. 

Multiple sclerosis (MS) is a disease of the nervous system in which myelin, the insulating substance that surrounds nerve cell axons, is damaged by the body's immune system \cite{MS,msclinical}. This damage causes individuals with MS to have neurological symptoms such as difficulty with coordination, vision impairment, pain, and fatigue, which can manifest acutely in episodic periods or lead to progressive decline of neurological function \cite{msclinical}.
Genome-wide association studies have identified some genetic risk-factors associated with MS and highlighted the role of the immune system in the disease but to the best of our knowledge, no one has used transcriptomic data and modeling to study the association of transcriptomic markers with MS \cite{gwas,gwas2,gwashafler}. 
Lack of molecular mechanistic understanding of the pathology of MS caused by an aberrant immune system complicates the ability in diagnosing and characterizing MS; indeed, there is no single test to diagnose MS \cite{tx}. Few molecular MS markers are known \cite{markers}. However, current indicators of MS are related to general neurological decline rather than the root causes of MS \cite{markers,tx}. Given the central role of the immune system in MS, we relied on a dataset composed of cerebrospinal fluid (CSF) and blood, both of which contain a high-proportion of immune cells. The selection of an immune-rich single-cell sample is critical to identifying molecular markers that can used to predict MS disease-state and to aid in developing a diagnostic test based on single-cell transcriptomic technology.

The existence of disease-modifying treatments highlights the value of this information-rich scRNA-seq data, as it can be used to characterize personalized disease-states, which may be differentially targeted with personalized treatment \cite{tx}. 
Current diagnoses of MS relies on imaging, neurological symptoms observed by a clinician and differential diagnoses, and spinal fluid abnormalities \cite{MS}. Currently, personalized approaches using a genetic risk score, based on genome-wide association studies, have poor agreement with clinical and MRI measures \cite{gwas,tx,badgwas}.

\begin{figure}[h]
  \centering
  \includegraphics[width=\linewidth]{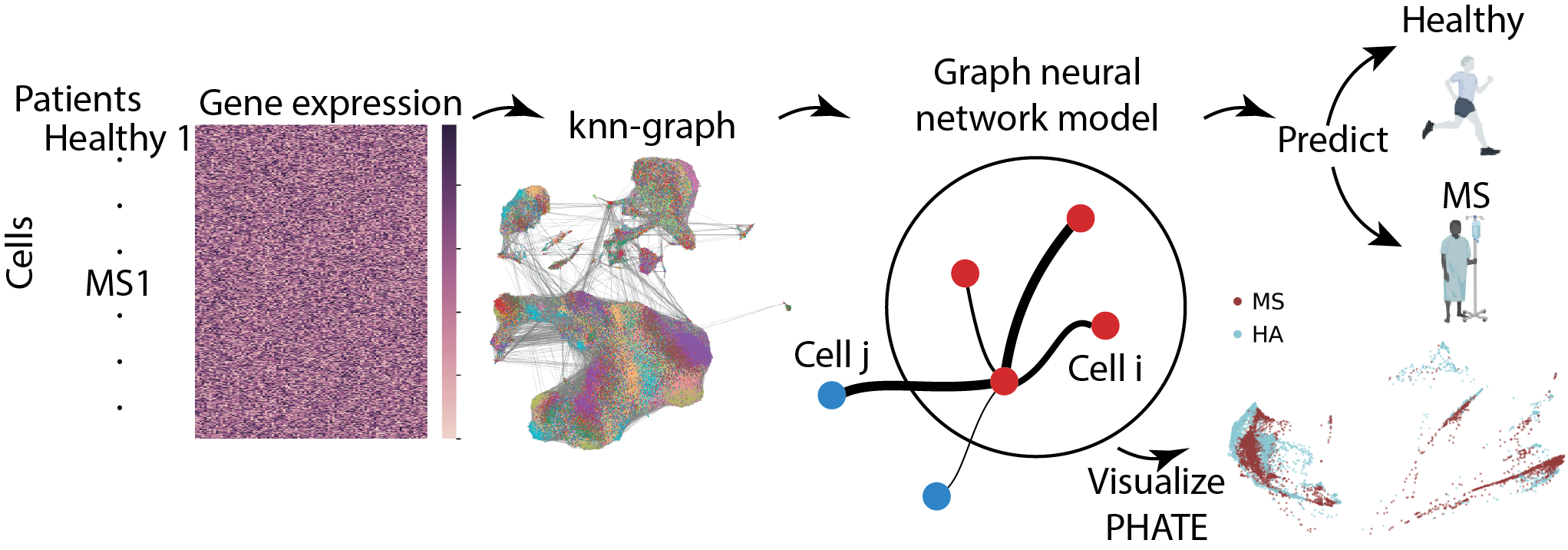}
  \caption{Schematic of our method depicting the use of a cell by gene count matrix in a graph attention model to predict and visualize disease state.}
  \Description{Workflow}
  \label{fig:model}
\end{figure}

Several classes of bio-molecular networks like transcriptional regulatory networks, protein-protein interaction networks, and metabolic networks are modeled as graphs. Mathematical graph theory is a straightforward way to represent this information, and graph-based models can exploit the global and local characteristics of these networks relevant to cell biology. In recent years, Graph Neural Networks (GNNs) have been widely adopted in various tasks, such as graph classification~\cite{convmol, attpool}, link prediction~\cite{linkpred} and node classification~\cite{GCN, GAT, Yang}. Various GNNs have been proven to be effective in achieving state-of-the-art performance in a variety of graph datasets such as social networks~\cite{Yang}, citation networks~\cite{GAT}, recommender systems~\cite{GCN, Yang} and protein-protein interaction networks~\cite{GAT}. The underlying graph structure is utilized by GNNs to operate convolution directly on graphs by passing node features to neighbors~\cite{Yang}, or to perform convolution in the spectral domain by using the eigenbasis of the Graph Laplacian operator~\cite{GCN}. Since the introduction of the self-attention mechanism in the influential work~\cite{Attention}, it has become extremely common to use self-attention for various sequence-sequence tasks like machine translations and learning representations. Self-attention allows the model to deal with variable sized inputs and focus on the most important part of the inputs. The self-attention mechanism has been introduced to the graph domains by \cite{GAT} and also developed by \cite{Bottou2005GraphTN, GRAM, GTN}. In our work, we chose to utilize the Graph Attention Network (GAT) as introduced by \cite{GAT} to do \textit{node classification} because of it's high performance on complex datasets, its adaptability to unseen datasets, and the model's interpretability. Figure~\ref{fig:model} shows the workflow of our paper. 

Our contributions are as follows: (i) we use graph neural networks on single cell data to predict disease state of MS patients and healthy individuals, (ii) we use the attention weights matrix to visualize the new graphs learned by the model, (iii) we extract meaningful features from the model and investigate them in the context of existing biological knowledge on MS.

\section{Datasets used}

\begin{figure}[h]
  \centering
  \includegraphics[width=\linewidth]{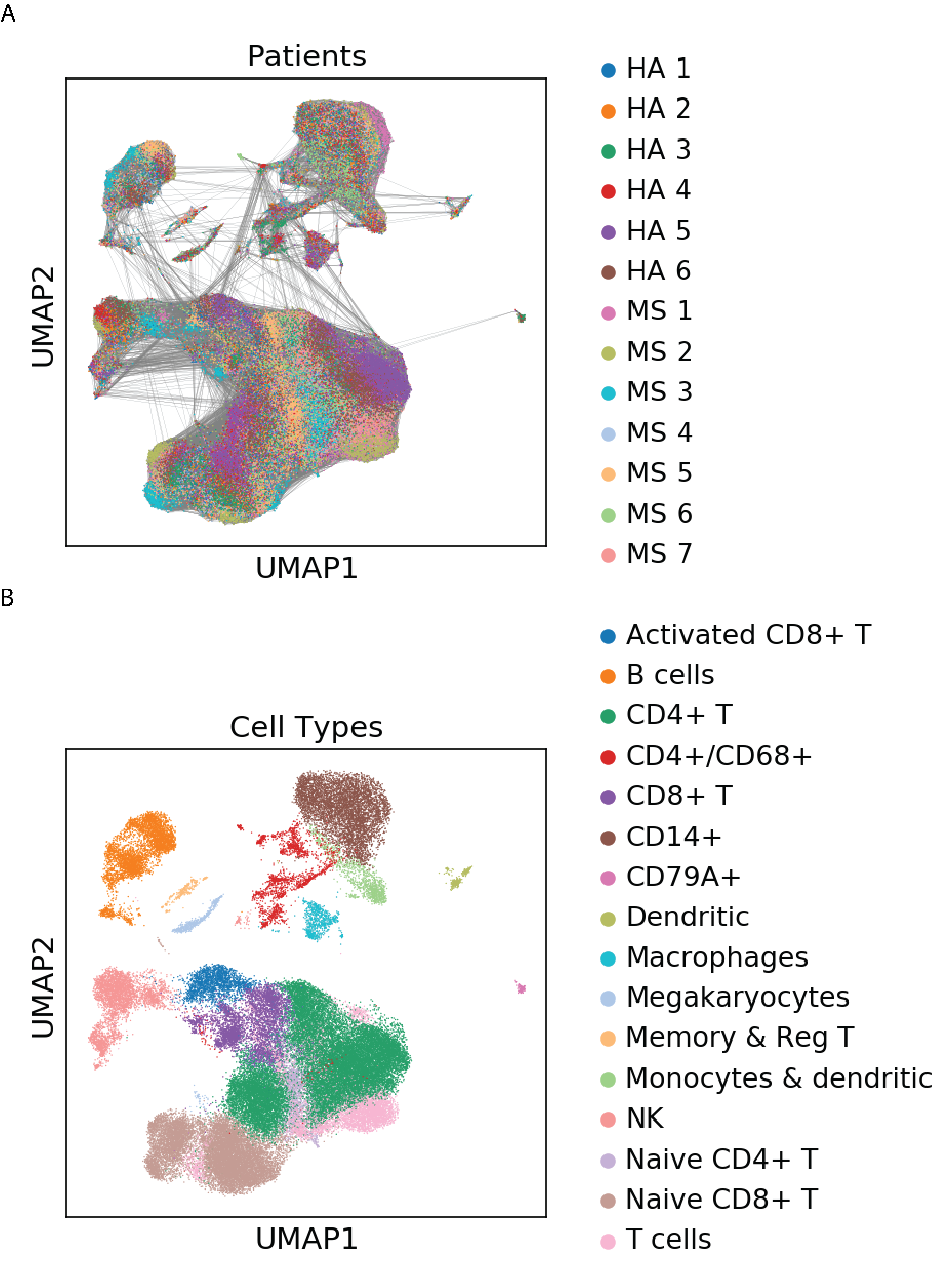}
  \caption{Low-dimensional embeddings of single-cell data used in this paper. A) UMAP scatter plot of the single-cell data used for training our model colored by the 6 healthy individuals (HA*) and 7 MS patients (MS*). Gray lines show edges in a nearest-neighbor graph. B) UMAP scatter plot colored by cell types identified in the blood and CSF.}
  \Description{Various cell types in our dataset}
  \label{fig:celltypes}
\end{figure}

We used a pooled dataset composed of blood and cerebrospinal fluid (CSF) samples from $7$ MS patients and $6$ healthy adults (HA), constituting $26$ scRNA-seq samples with an average of $2,500$ cells per sample and about $5,000$ cells per patient. We performed standard scRNA-seq quality-control filtering steps to the raw data, including removing cells expressing $>10$\% mitochondrial genes, normalizing gene counts by the sequenced library size per cell, square-root transforming gene counts, and removing mitochondrial and ribosomal protein transcripts from downstream analysis \cite{Scanpy}. After the preprocessing steps our dataset is composed of $66,667$ cells with $22,005$ gene features per cell. Figure~\ref{fig:celltypes}(A) shows our full graph (used in \textit{transductive} task) colored by the individual patients and figure~\ref{fig:celltypes}(B) is colored by various cell types. Variability between samples dominates biological variability in reduced dimensional embeddings and these batch effects complicate model inference of biological features \cite{batch}. We treat each cell as a node, so our dataset has $22,005$ features per node, with high variability across features. The number of features is an order of magnitude higher than the citation, social-network, and protein-protein interaction datasets previously used by GNNs. This feature matrix was used to compute a k-nearest neighbor graph from euclidean distances in PCA reduced space ($10$ neighbors and $100$ principal components) and clustered by the Louvain algorithm to identify cell types within the pooled dataset~\cite{2018arXivUMAP,louvain}. This led to the unsupervised identification of at least $16$ cell types within the pooled dataset, including about $75$\% T cells broken into various lineages and about $15$\% B cells, in addition to other immune and peripheral blood mononuclear cells, including macrophages, monocytes, natural-killer cells, and platelets. The nodes are classified as healthy or MS. Thus the classification problem we are interested in is at the \textit{cell level}.

We use this dataset to perform two tasks: (i) inductive inference, and (ii) transductive inference.  For the \textit{transduction} task, we randomly assign $10\%$ of the nodes for validation and $10\%$ for testing while keeping the ratio of healthy and MS cells the same as in the full dataset. Then we randomly mask the labels of $50\%$ of the training, validation and the test set. Thus we can think of our problem as having a large graph where half of the graph is unlabeled and our goal is to predict the labels based on the labels of the nodes in the graph and the features of the node. With the transductive setup in mind, we feed in the features of \textit{all} nodes during training and our goal is to predict the masked labels.


For the \textit{induction} task, we randomly choose a healthy adult and a MS patient each for the validation and test sets and the remaining $5$ MS patients and $4$ healthy adults for training. We construct separate knn-graphs for the training, validation, and test sets just like before. It is important to note that the testing graphs are completely unseen during training. Table~\ref{tab:single_cell_dataset} gives a complete description of the datasets used in various experiments. 

\begin{table}[h!]
  \caption{Characteristics of our single cell dataset}
  \label{tab:single_cell_dataset}
  \centering
  \begin{tabular}{|c|c|c|c|c|}
    \hline
   Task &  & Train & Dev & Test\\
    \hline
   Inductive & \# Nodes & $43866$ & $9686$ & $13033$\\
   & \# Edges & $332398$ & $73552$ & $100715$ \\
 & \# Features & $22005$ & $22005$ & $22005$\\
 & \# Classes & $2$ & $2$ & $2$ \\
 & \# Graphs & $1$ & $1$ & $1$ \\

  \hline
  
 Transductive & \# Nodes & $54000$ & $6000$ & $6667$\\
    & \# Features & $22005$ & $22005$ & $22005$ \\
     & \# Classes & $2$ & $2$ & $2$ \\ \cline{2-5}
      & \multicolumn{3}{|c}{\# Edges \qquad  $\> 5007093$} &\\ 
    \hline
\end{tabular}
\end{table}

\subsection{Related work}
Machine learning, and particularly deep learning have been widely used in the medical domain to predict diseases but mostly using medical images or EHR data. Machine learning has also been used on scRNA-seq data for data de-noising, batch correction, data imputation to correct for sparse signal detection or dropout, unsupervised clustering, cell-type prediction \cite{rnacluster,denoising,imputation,cellpred} and low-dimensional visualization of samples relevant to human health \cite{saucie,denoising}. Recently, transfer learning (using an autoencoder) has been used to extract new gene-gene interactions~\cite{rnatransfer} and random walks have been used to predict gene responses to drugs~\cite{Harikumar837807}. In the biomedical domain GNNs have mostly been used for prediction of protein-protein interactions~\cite{GAT} and gene-gene interactions\cite{graphreview, Harikumar837807}. However, to the best of our knowledge, predicting diseases from single-cell RNA seq data using state-of-the-art methods have not been tried successfully.

\section{Architecture of our model}

We use the Graph Attention Network by \cite{GAT} and we follow the exposition in \cite{GAT}. In this section we describe our model in detail. The building block of our network is the graph attention layer. The input to our layer is a set of node features, $\mathbf{h} = \{ h_1, h_2, . . . , h_N \}$, where $h_i \in \mathbb{R}^F $, $N$ is the number of nodes, and $F$ is the number of features in each node. The layer produces a new set of node features (of possibly different cardinality $F'$) as its output, $\mathbf{h'} = \{h'_1,h'_2,....h'_N \}$ where $h'_i \in \mathbb{R}^{F'}$.

 We apply a linear transformation called the weight matrix, $\mathbb{W} \in \mathbb{R}^{F'\times F}$ to every node. The weight matrix $\mathbb{W}$ is initialized with Glorot initialization~\cite{Glorot}. Self-attention is then computed on the nodes, i.e. a shared attention mechanism 
 \begin{equation}
     a : \mathbb{R}^{F'} \times \mathbb{R}^{F'} \rightarrow \mathbb{R}
 \end{equation}
that computes attention coefficients
\begin{equation} 
e_{ij} = a(\mathbb{W}h_{i}, \mathbb{W}h_j )
\end{equation}
We treat these $e_{ij}$'s as importance scores of node $j$’s features to that of node $i$. This is the main difference between a graph attention network and a graph convolution network \cite{GCN}. Unlike \cite{GCN}, different importance scores are given to different nodes in the same neighborhood, which allows for greater adaptability of the model to various complicated datasets like ours. 

Furthermore, we normalize these $e_{ij}$'s across all choices of $j$, where $j$ lies in some neighborhood of $i$ using the softmax function:
\begin{equation}\label{eqn:att_coeff}
\alpha_{ij} = \text{softmax}_{j} (e_{ij} ) = \frac{\text{exp}(e_{ij} )} {\sum_ {k \in \mathcal{N}_{i}} \text{exp}(e_{ik})}
\end{equation}
where $\mathcal{N}_i$ is some neighborhood of the node $i$. In all our applications, the graphs are the KNN graphs and $\mathcal{N}_i$ is the first order $k$-neighbors of the node $i$. Thus in practice, we only compute $e_{ij}$ (and thus $\alpha_{ij}$) for all nodes $j$ in the first-order neighborhood of the node $i$, instead of all pairs of nodes $i$ and $j$.  

In our experiments, the attention mechanism $a$ is a single-layer feedforward neural network on which we apply the LeakyReLU non-linearity. The normalized attention coefficients are then used to compute a linear combination of the features corresponding to them, to serve as the final output features for every node (after applying a nonlinearity, $\sigma$):
\begin{equation} \label{eqn:new_feat}
h'_{i} = \sigma \bigg( \sum_{j \in \mathcal{N}_i} \alpha_{ij}\mathbb{W}h_{j} \bigg).
\end{equation}

As is now common in attention based networks, we use multi-head attention, similar to \cite{Attention, GAT}. In this case, each attention head learns an unique set of attention weights independent of the other heads in a given layer. Specifically, $K$ independent attention mechanisms execute the transformation of Equation $4$, and then their features are concatenated, resulting in the following output feature representation:

\begin{equation} 
h'_{i} = \bigg|\bigg|_{l=1}^{K} \sigma \bigg( \sum_{j \in \mathcal{N}_i} \alpha^{l}_{ij}\mathbb{W}^{l}h_{j} \bigg).
\end{equation}
where $||$ denotes the concatenation operation, $\alpha^{l}_{ij}$ are normalized attention coefficients computed by the $l$-th
attention mechanism and $\mathbb{W}^{l}$ is the corresponding input linear transformation’s weight matrix.
Note that, in the final output, each node will have $KF'$ features.\\

Finally, in the final (prediction) layer of the network, we first employ averaging, and then apply the final nonlinearity (logistic sigmoid for our classification problems). Thus the equation for the final layer is: 
\begin{equation}
    h'_{i} = \sigma \bigg(\frac{1}{K}\sum_{l=1}^{K} \sum_{j \in \mathcal{N}_i} \alpha^{l}_{ij}\mathbb{W}^{l}h_{j} \bigg).
\end{equation}

We also use weight clipping to prevent the gradients from exploding. Table~\ref{tab:hyperparameters} shows all the hyperparameters in our models used for the experiments. Our code is written in PyTorch and is publicly available at \url{https://github.com/vandijklab/scGAT}.

Another reason of working with Graph Attention Networks is it's power of interpretability. It is easy to visualize the attention heads, i.e. the attention scores between various nodes by a head in any given layer. Due to memory issues, we implemented a sparse version of the code of \cite{GAT}. 

\section{Experiments}

In this section we will explain in detail our experiments. We compare our work with 
random forests, a multilayer perceptrons, Graph Convolution Networks as introduced in~\cite{GCN}, and a dummy estimator. Table~\ref{tab:exp_results} gives complete details of our results for the following tasks. All reported numbers are from the test sets. We also experimented with changing the default hyperparameters like changing the width and breadth of the model, the learning rate, optimizers, losses, and the dropout rate. However the best results were obtained by our default hyperparameters.

Generally all deep learning models are trained by stochastic gradient descent which requires us to minibatch our graphs. To breakdown a large graph for memory purposes and fast efficient training, we follow the algorithm in \cite{clusterloader} which is implemented in PyTorch Geometric. For the \textit{inductive} task, we break up our training graph into $4000$ roughly equal subgraphs and use a batch size of $256$ graphs. The average number of nodes in each batch is $2741$. But we do not split the the test or the validation graphs. For the \textit{transductive} task, we break our graph as before in $4000$ subgraphs and use a batch size of $256$ subgraphs. The average number of nodes per subgraph in a batch is $4657$.

 \begin{table}[h!]
\centering
\caption{Default hyperparameters used in the experiments}
\resizebox{\columnwidth}{!}
{
\begin{tabular}{|c|c|c|}
\hline
 & Graph Attention Network & Graph Convolution Network \\
\hline
Number of layers & $2$ & $2$ \\
\hline
Hidden\_size & $8$ & $256$ \\
\hline
Attention Heads & $8$ & N/A \\
\hline
Optimizer & Adagrad & Adagrad\\
\hline
clip\_grad\_norm & $5$ & $2$\\
\hline
learning rate scheduler & cosine\_decay \& $1$st decay steps = $1000$ & cosine\_decay \& $1$st decay steps = $1000$\\
\hline
weight\_decay & $.0005$ & $.005$ \\
\hline
Batch size & $256$ & $256$\\
\hline
Dropout & $.5$ & $.4$ \\
\hline
Slope in LeakyRelu & $.2$ & $.2$ \\
\hline
Training Epochs & $1000$ & $1000$ \\
\hline
Early stopping &  $100$ & $100$\\
\hline
\end{tabular}

\label{tab:hyperparameters}
}
\end{table}

\subsection{Transductive learning} We use our single cell dataset and we follow the transductive learning setup as in \cite{Yang}. The transductive setup implies that during training the model has access to all the nodes' feature vectors. The predictive power of the model is then evaluated on the masked test nodes and we use masked validation nodes for validation purposes. Our main task is to predict if a cell is healthy or has MS. We ran our GAT model $8$ times.  

\begin{table}[h!]
\centering
\caption{Experimental results}
\label{tab:exp_results}

\resizebox{\columnwidth}{!}
{
\begin{tabular}{|c|c|c|}
\hline
Task & Model & Accuracy \\
\hline
 & Random & $51.8$ \\
Inductive & MLP & $56.7$ \\
 & Random Forest & $58.5$ \\
 & Graph Convolutional Network & $72.1$ \\
 & Graph Attention Network(our) & $\mathbf{92.3} \pm .7$\\
 \hline
 Transductive & Graph Convolutional Network & $82.91$ \\
    & Graph Attention Network(our) & $\mathbf{86} \pm  .3$\\
    \hline

\end{tabular}
}
\end{table}

\begin{figure*}[h]
  \centering
  \includegraphics[width=\textwidth]{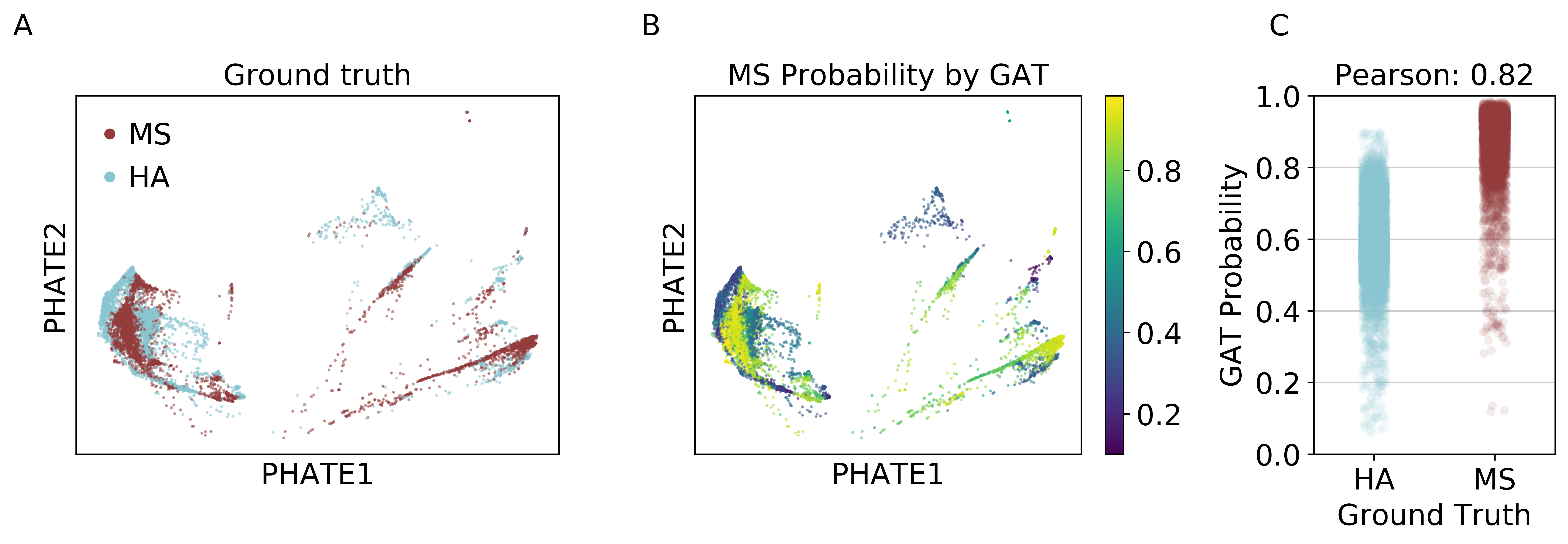}
  \caption{Predicted probabilities from induction task per cell. A) PHATE plot of the data colored by ground truth labels. B) PHATE plot colored by the predictions of our graph attention model. C) Model prediction probability (Y-axis) for indicated ground truth label (X-axis).}
  \Description{Visualizing MS predictions vs the ground truth via Phate.}
  \label{fig:pred_dist}
\end{figure*}

\subsection{Inductive learning} 
For this task we again use our single cell dataset. We have $13$ persons' blood and CSF data and out of them $6$ of them have MS and the rest are healthy. We randomly choose a blood and CSF data of a healthy adult and a MS patient each for the validation and test sets and the remaining $5$ MS patients and $4$ healthy adults for training. We construct separate knn-graphs for the training, validation, and the test sets. We ran our GAT model $16$ times. 

\begin{figure}[h!]
 \centering
  \includegraphics[width=\linewidth]{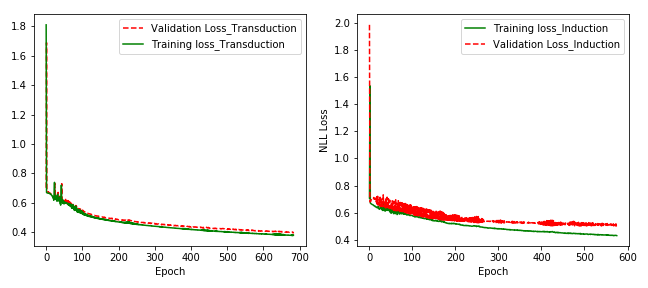}
  \caption{Negative log-likelihood losses for training and validation sets in the induction (left) and transduction (right) tasks.}
  \Description{Visualizing MS predictions vs the ground truth via PHATE.}
  \label{fig:loss}
\end{figure}

Figure~\ref{fig:loss} shows the loss of our model for the induction and the transduction tasks. 

\section{Discussion of our results}
Encouraged by the predictive ability of our model in both transductive and inductive tasks, we take a deeper look at the attention heads and attention weights at various layers of our model. In this section, we will extract meaningful biological information from the attention heads.

\subsection{Finding interesting gene markers} We look at $\mathbb{W}_{h}^T$ where $T$ stands for transpose and $\mathbb{W}_h$ stands for the learned weight matrix for head $h$ in layer $1$. $\mathbb{W}_h$ in our case is an $8 \times 22005$ matrix where $8$ is the size of the hidden dimension. For each row $i$ of each $\mathbb{W}_{h}^T$, we take the max of the absolute value and we define 
\begin{equation}
    g_{i}^h = \text{max}_{j}(|w_{ij}|)
\end{equation}
and then we pick the highest $10$  $g^{h}_{i}$ for each $h$. We can think of these genes as the most important features the model learns to discriminate between healthy and MS cells. 

\begin{figure}[h!]
  \centering
  \includegraphics[width=\linewidth]{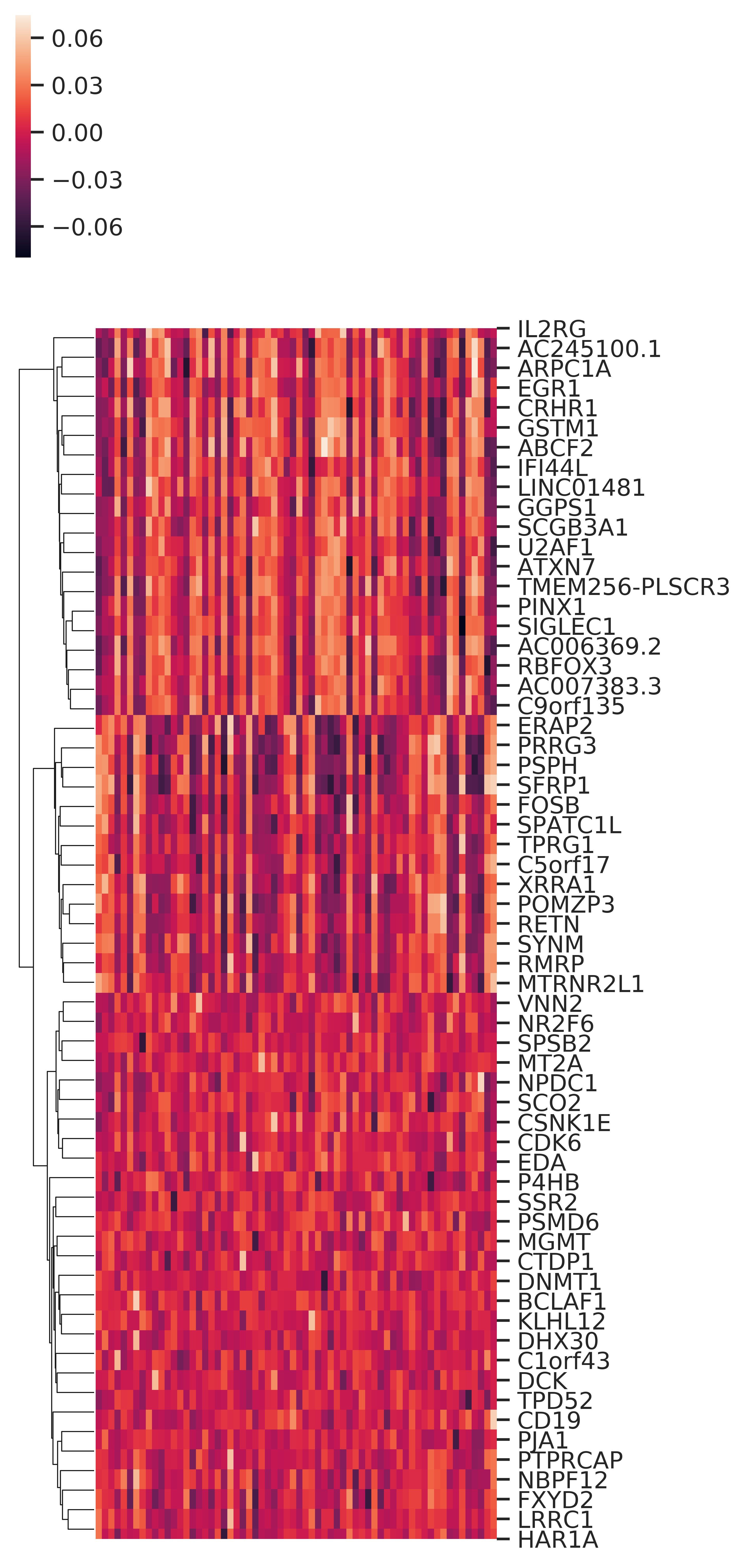}
  \caption{Heatmap showing the top 10 gene features for each head colored by learned weights in the induction task and clustered by gene.}
  \Description{Gene list.}
  \label{fig:gene_list}
\end{figure}

A heatmap of these features for the model trained in the inductive task is shown in figure~\ref{fig:gene_list}. We are encouraged to find \textbf{IL2RG} as a learned feature in discriminating between healthy and MS cells. The interleukin-2 receptor (IL2R) is involved in a signaling pathway that is essential for T cell function and genetic variants of IL-2R are associated with various autoimmune disorders, including MS \cite{il2r,gwashafler}. In addition, we are encouraged that the network gives a high weight to \textbf{CD19}, a marker for B cells, since B cells are targeted by therapeutics and is thought to contribute to MS by secreting autoreactive antibodies that cause neuroinflammation and demyelination \cite{Hauser2017}. Gene Ontology of the other identified features revealed that $38$ out of $58$ mapped genes are involved in the regulation of the hormone secretion while a number of other genes in our list are involved in the regulation of the nerve cell development. Lipid metabolism and nerve cell development are critical to normal myelination, so alterations to these processes may lead to a triggering of the autoimmune system and cause inflammation related to development of MS pathologies \cite{MS,mspersonalized}. 

Thus it is encouraging to see that representations learned by our model align with the medical knowledge. Given the paucity of molecular markers that can characterize MS disease states, other than genes identified through genome-wide association studies and metabolites indicative of neurological degeneration, we presume that our model and this novel list of gene features will guide further experimental work into the pathophysiology of MS \cite{tx}. Broadly, this feature selection approach promises to reveal biomarkers of MS disease pathologies, which may be used for more cost-effective diagnostic testing than the single-cell technologies affords like DNA microarray screening or measuring gene expression by flow cytometry \cite{mspersonalized}. 

\subsection{Finding cell types most prototypical of MS}

In both the tasks, we computed the top $100$ cells that have the highest predicted probability of being MS. These are the cells that are strongly indicative of MS. We noted that all of these cells are from patients' CSF. In contrast, the $100$ most prototypical healthy cells are approximately evenly split between being sourced from blood or CSF samples. This confirms the common medical hypothesis that patients' CSF has a higher degree of information relevant to central nervous system inflammation than blood cell samples \cite{Lycke2017}. In addition, these observations support that single-cell technology may supplement existing neuroinflammation diagnostic criteria relying on detection of spinal fluid abnormalities \cite{Lycke2017,tx}. 

\begin{figure}[h!]
  \centering
  \includegraphics[width=\linewidth]{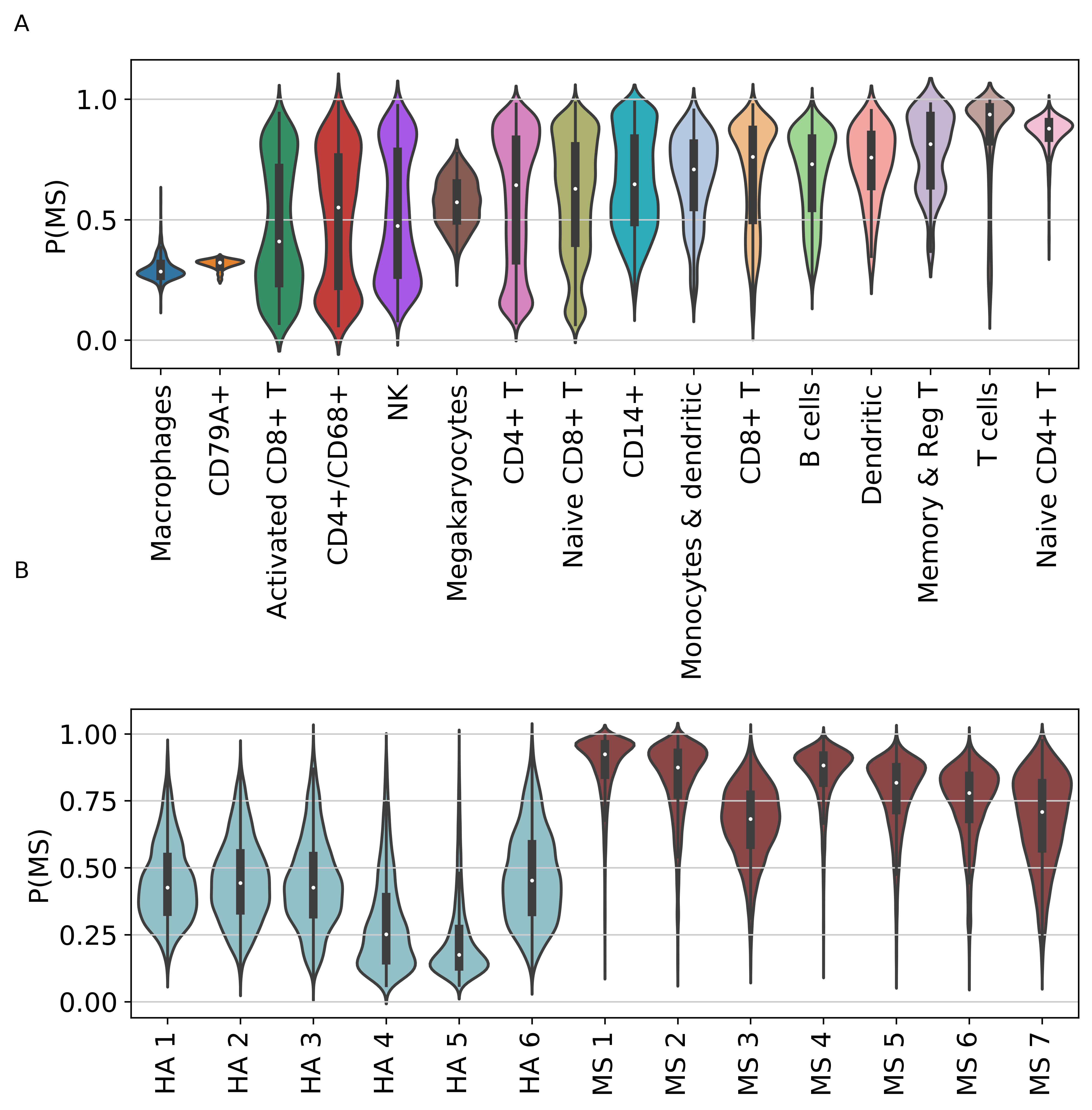}
  \caption{Predicted probabilities from transduction task for MS class label stratified by cell type and individual patient. Violin plot shows kernel density estimate and inner box plot shows interquartile range and median. A) Probabilities of having MS predicted by our model for each cell type. B) Probabilities of having MS predicted by our model for each patient.}
  \Description{Cell type preds.}
  \label{fig:trans_ctype}
\end{figure}

We assess the predicted probability of being MS, stratified by cell type in figure~\ref{fig:trans_ctype}A. Cells involved in innate immunity, such as macrophages and natural-killer cells, have a relatively lower average predicted probability of being MS compared to cells involved in adaptive immunity, such as B cells and T cells. This reflects the current medical understanding of the immune system's role in MS and the key role that adaptive immunity plays in autoimmune disorders and neuroinflammation more broadly \cite{tx,msimmune1,msimmune2}. This suggests that our model is tuned to recapitulate biological knowledge. 

We assess the predicted probability of being MS, stratified by patient in figure~\ref{fig:trans_ctype}B. MS is known to present in different forms and the severity of the disease can vary across patients \cite{msclinical}. Our model performs a node classification task, yielding the predicted probability of being MS for every cell drawn from a single patient. Thus, a single patient can have cells that are more prototypically healthy than MS. The average predicted probability is as low as $0.23$ for healthy patients and as high as $0.88$ for MS patients in the transduction task. Both the transduction and induction tasks have models that achieve good discriminability between disease-states, with a Pearson's correlation coefficient of $\mathbf{0.76}$ and $\mathbf{0.82}$, respectively.

\subsection{Visualizing Attention heads}
In this subsection, we will take a deeper look at the features that our GAT model learned and visualize the latent space using \cite{2018arXivUMAP} and \cite{Scanpy}. We can construct a matrix whose $(i,j)$ entry is $\alpha_{ij}$ as defined by equation \ref{eqn:att_coeff} for each head in the first layer. We can think of these attention matrices as normalized adjacency matrix of a directed graph (i.e. the stochastic matrix one obtains from normalizing the adjacency matrix of a graph). 

\begin{figure}[h!]
  \centering
  \includegraphics[width= .8\columnwidth]{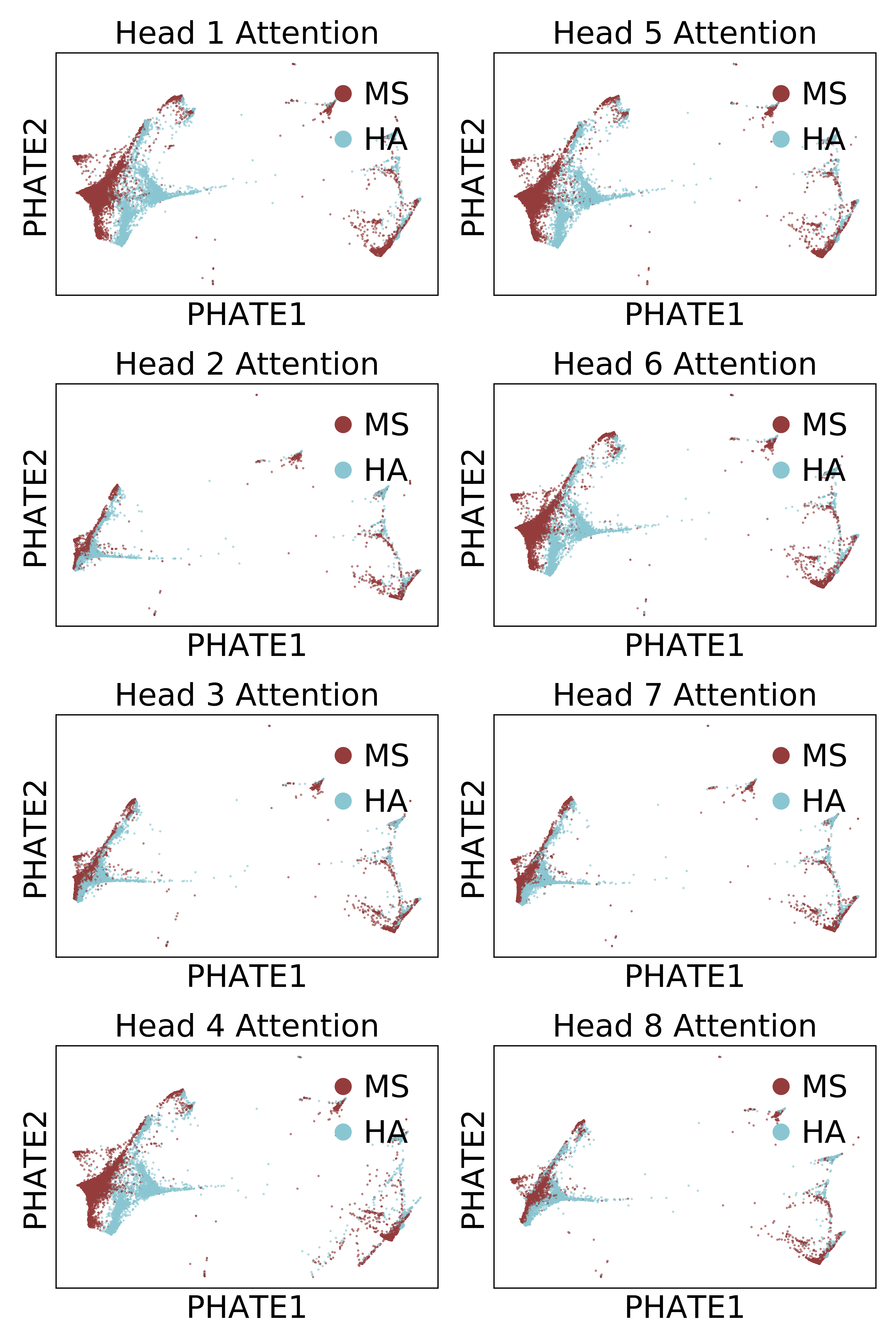}
  \caption{New low-dimensional embeddings learned by the PHATE algorithm after inputting graphs learned by various attention heads in our model for the induction task.}
  \Description{Visualizing learned graphs.}
  \label{fig:phate_att}
\end{figure}

\begin{figure}[h!]
  \centering
  \includegraphics[width=\linewidth]{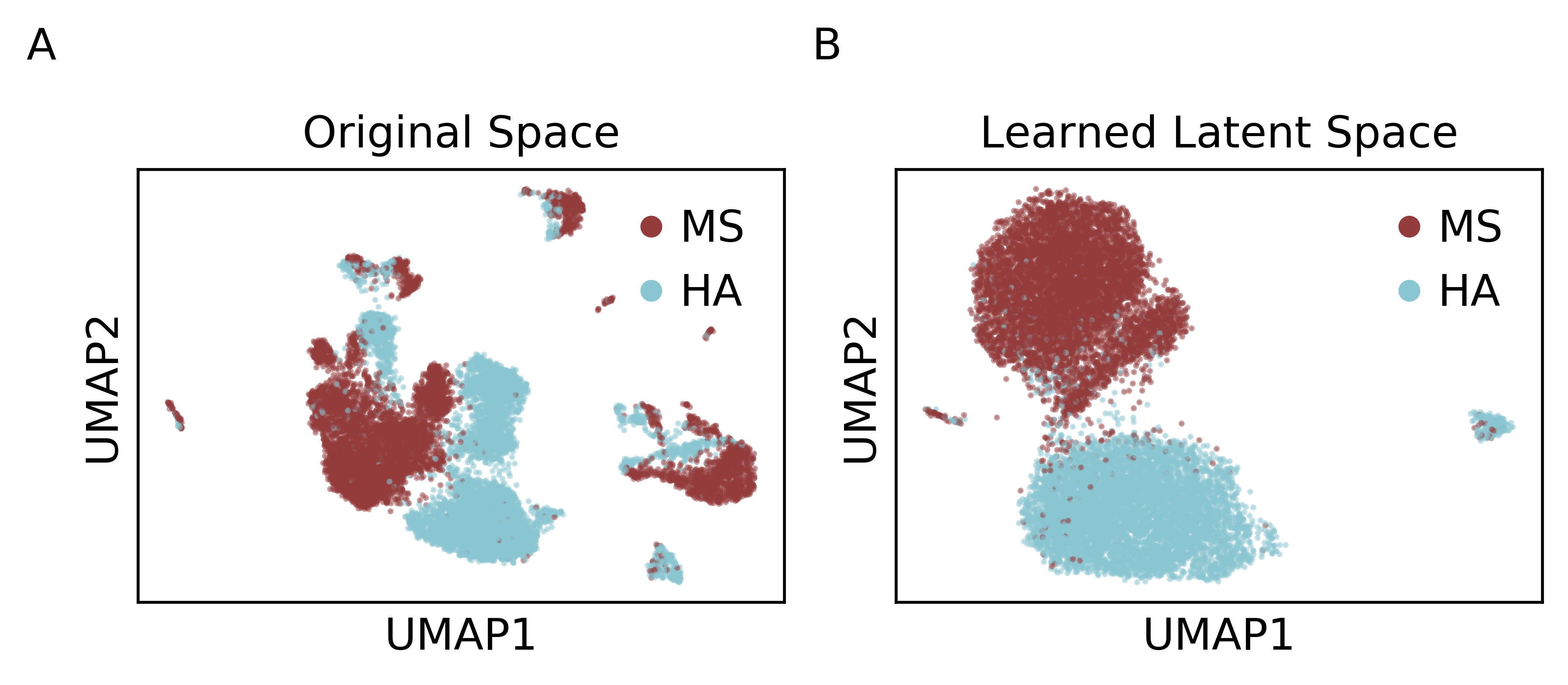}
  \caption{Visualization of the latent space learned by our Graph Attention Model during the induction task. The UMAP plot of the original graph (left) and the UMAP plot of the latent space (right).}
  \Description{Visualizing combined heads.}
  \label{fig:combined_head}
\end{figure}

And we can feed it into PHATE (figure~\ref{fig:phate_att}), allowing us to visualize the new graphs that our model has learned. We can easily see that head $4$ and head $8$ learn very different graphs. On the other hand, we can look at the learned $\mathbf{h'}$ as in equation~\ref{eqn:new_feat} by each head in each layer which we view as the new features or the latent space. Just as before, we can feed these new features into a manifold learning algorithm and can construct new KNN graphs (with $10$ neighbors) for visualization with UMAP, as shown in the figure~\ref{fig:combined_head} for a $64$ dimensional feature space constructed by concatenating the hidden units for each each head per node. Here we see that the  learned latent space separates healthy and MS cells in the low-dimensional embedding. This further supports that our model learns to discriminate between healthy and MS labels.

\section{Future work}

We are excited by the improvements that a Graph Attention Networks bring to single cell analysis. We would like to continue this direction further by incorporating in multidimensional edge level features. Another direction of our research is to come up with new semi/unsupervised ways of creating new graphs from single cell data by completely bypassing PCA and KNN graphs. And then we can apply the models in this paper along with the edge level modifications to our new graphs. \\

We are going to scale up our work to a much larger dataset which requires us not only to experiment with the depth and the breadth of our model but also introduces new computational challenges like fitting large graphs in memory and doing computations more effectively. \\

One of the striking results we found that conforms to the medical knowledge is that the CSF cells are strong indicators of MS, whereas the blood cell are not very good indicators. However getting the CSF data is an expensive and invasive process. We are in process of acquiring only the blood samples of MS patients. Thus predicting MS from only blood samples becomes an extremely difficult task. To alleviate the problem, we would like to do \textbf{transfer learning} from our model that is trained on both blood and CSF samples and the train/test \textbf{only} on the new blood samples. We hope that the transfer learning will achieve better results than an untrained graph neural network.

\section{Conclusion}
Multiple sclerosis (MS) is an unpredictable disease of the central nervous system that disrupts the flow of information within the brain,
and between the brain and body. The molecular markers and characteristics of MS are still actively studied and there remains no single test for diagnosing MS. Single-cell sequencing has revolutionized biological discovery, providing an in-depth picture
of cellular differences and the interactions between various cells in their microenvironments. However, scRNA-seq has mostly been used for basic sciences but not for clinical or diagnostic use due to its
cost and the complexity of the data. In this work we use a graph attention model on single cell data to predict MS. The model is trained on a cohort of seven MS patients
and six healthy adults (HA) from the blood and cerebral spinal fluid
(CSF) of each patient. We perform transductive and inductive tasks and we achieve $\mathbf{86}$ \% and $\mathbf{92}$ \% accuracy in predicting MS beating other methods such as graph convolutional networks, random forest and MLP. Further, we use the learned graph attention model to get insight into the signatures (transcriptomic and genetic features) and the cells that are important for this prediction. Using the attention weights we extracted a list of $100$ genes and we were encouraged to find \textbf{CD19} and \textbf{IL2RG} in that list. We envision that the feature selection approach to produce a minimal gene set can function as biomarkers in a clinical setting. We also computed the top $100$ cells that have the highest predicted probability of being MS and all of these cells came from the patients' CSF. In contrast, the $100$ most prototypical healthy cells are approximately evenly split between being sourced from blood or CSF samples. This confirms the common medical hypothesis that patients' CSF has a higher degree of information relevant to central nervous system inflammation than blood cell samples. Finally the graph attention model also allow us to infer a new feature space for the cells that emphasizes the difference between the two conditions which we visualize via PHATE and UMAP. There has been more emphasis on the identification of determinants of disease progression and on how individual information can be used to  personalize treatment and we hope that this graph neural networks will play in important role in predicting disease and in identifying key determinant of those diseases.

\bibliographystyle{ACM-Reference-Format}
\bibliography{ms_ref}
\nocite{*}

\end{document}